\documentclass[prl,twocolumn]{revtex4}
\usepackage{graphicx}
\usepackage{dcolumn}
\usepackage{color}
\usepackage{latexsym}
\usepackage{bm}
\usepackage{amsmath}
\usepackage{amsfonts}   
\usepackage{amssymb}
\usepackage{multirow}
\begin{document}
%\date{}
%%%%%%%%%%%%%%%%%%%%%%%%%%%%%%%%%%%%%%%%%%%%%%%%%%%%%%%%%%%%%%%%%%%%%%%%%%%

\title{New approach to phase transitions in black holes}
\author{Rabin Banerjee}
\email{rabin@bose.res.in}
\author{Sujoy Kumar Modak}
\email{sujoy@bose.res.in}
\affiliation{
S. N. Bose National Centre for Basic Sciences, Block-JD, 
Sector-III, Salt Lake, Kolkata 700 098, India
}
\author{Saurav Samanta}
\email {srvsmnt@gmail.com}
\affiliation{Narasinha Dutt College, 129, Belilious Road, Howrah-711101, India}
%%%%%%%%%%%%%%%%%%%%%%%%%%%%%%%%%%%%%%%%%%%%%%%%%%%%%%%%%%%%%%
\begin{abstract}
We develop an analogy between fluids and black holes to study phase transitions in the latter. The entropy-temperature graph shows the onset of a phase transition without any latent heat. The nature of this continuous (higher order) phase transition is examined in details. We find that the second order derivatives of the free energy diverge at the critical temperature. Also, the transition is smeared instead of sharp, so that the usual Ehrenfest's scheme breaks down. A generalised version of this scheme is formulated which is shown to be consistent with the phase transition curves.
\end{abstract}

\maketitle
Although black holes are well known thermodynamical systems, they are not well understood. In fact there is no microscopic or statistical description behind their thermodynamical behaviour. Studying phase transition in black holes is very important as it could manifest the underlying micro-structure of black hole thermodynamics. Although there are some investigations \cite{pt}, these are more concerned with the geometrical rather than the thermodynamical aspects of phase transition.

In this paper we systematically develop a new approach to study phase transitions in black holes from a thermodynamical viewpoint. To do that we first build an analogy between fluids and black holes that parallels the analogy between between fluids and magnetic systems. From this analogy we then exploit the known results in fluids to infer the corresponding situations in black holes. Following this approach we find that the entropy-temperature graph (as well as the Gibb's energy-temperature graph) of the Kerr-AdS black hole shows the onset of phase transition without any latent heat. Moreover this transition is smeared round the critical temperature. The details of this continuous (higher order) phase transition are examined by looking at various plots of entropy with specific heat, volume expansion and compressibility. All these plots manifest an infinite divergence at the critical point. Furthermore, the smeared nature of the phase transition gets highlighted.

Due to this smeared transition a direct application of the Gibbsian approach (Ehrenfest's scheme) to exhibit and classify phase transitions fails. We thus develop a generalized Ehrenfest scheme to account for this smearing. We show that such a generalization is compatible with our graphical analysis.       

The discovery that black holes are indeed thermodynamical systems is built on a mathematical analogy between laws of black hole mechanics and laws of thermodynamics \cite{bardeen}. For chargeless, rotating black holes the first law of black hole mechanics relates the infinitesimal change in black hole mass ($M$) with the infinitesimal changes in its horizon area ($A$) and angular momentum ($J$), given by
\begin{eqnarray}
\delta M= \frac{\kappa}{8\pi}\delta A+ \Omega\delta J,\label{bhm}
\end{eqnarray} 
where $\kappa$ and $\Omega$ are surface gravity and angular velocity at the event horizon, respectively. This is very much analogous to the first law of thermodynamics,
\begin{eqnarray}
dE= TdS-PdV \label{fl}.
\end{eqnarray} 
The fact that the internal energy ($E$) of black holes being represented by its mass ($M$), together with Hawking's discovery that black holes have temperature $T=\frac{\kappa}{2\pi}$, set the analogy between (\ref{bhm}) and (\ref{fl}), associating an entropy to black holes $S=\frac{A}{4}$. Several alternative approaches \cite{other} have been developed which reproduce identical results for black hole temperature and entropy.  

We observe that while the first terms in the right hand sides of (\ref{bhm}) and (\ref{fl}) set a mathematical analogy between black holes and ideal fluids, it is the second terms which actually establish the physical motivation. It leads to the correspondence $P\longleftrightarrow -\Omega$ and $V\longleftrightarrow J$. One can realise the physical importance of this analogy in the following manner. In ideal fluids, both for isothermal and isentropic processes if one increases the pressure its volume shrinks so that the density increases. Thus the $P-V$ plot for fluids has a negative slope. On the other hand the relation between the angular velocity and angular momentum of Kerr-AdS black hole is given by \cite{Cald,modak} 
\begin{eqnarray}
J=\frac{S \sqrt{\frac{S \Omega ^2}{(\pi +S) \left(\pi +S-S \Omega ^2\right)}} \left((\pi +S)^2-S^2 \Omega ^2\right)^2}{2 \pi ^{3/2} (\pi +S)^3}.
\label{JJ}
\end{eqnarray}
We plot the $\Omega-J$ graph in figure-1 by taking various constant values for entropy.
\begin{figure}[h]
\centering
%\rotatebox{270}{
\includegraphics[angle=0,width=6cm,height=3cm]{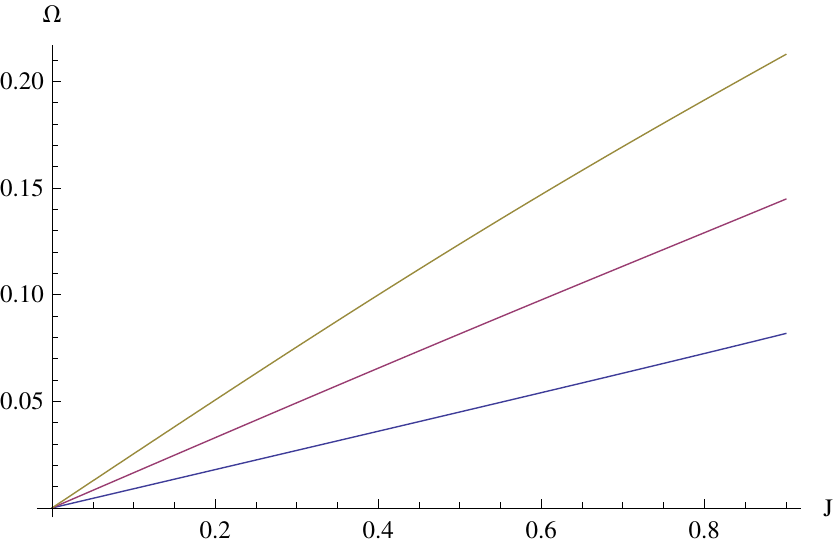}
\caption[]{Angular velocity ($\Omega$) vs. angular momentum ($J$) plots. For blue line (lower )$S=1$; for purple line (middle) $S=3$; and for grey line (upper) $S=5$.}
\label{figure 1a}
\end{figure}
All these plots have positive slopes which give a physical credibility to the correspondance,
%\begin{table}[th]
%\begin{tabular} [c] {|c|c|}\hline 
%\multirow{2}{*}{\bf x} & \bf $\rho_0$($\mu\Omega$ & \bf A($\mu\Omega$ & \bf B$\times 10^{-4}$($\mu\Omega$ & \bf$\Theta_{D}$ & \bf C$\times 10^{-2}$($\mu\Omega$ \\ 
       % & \bf $-cm$) & \bf $-cm$) & \bf $-cm/K^{3/2}$) & \bf (K) & \bf $-cm/K^{1/2}$) \\ \hline
  %Fluids & Black Holes\\ \hline\hline      
  %Energry ($E$) & Mass ($M$)\\ \hline
  %  Pressure ($P$) & Angular velocity ($-\Omega$)\\ \hline
  %  Volume ($V$) & Angular momentum ($J$)\\ \hline
  %  Temperature ($T$) & Temperature ($T$)\\ \hline
  %  Entropy ($S$) & Entropy ($S$)\\ \hline 
%\end{tabular}
%\caption {Comparison between ideal fluids and black holes}
%\end{table}
\begin{eqnarray}
E\longleftrightarrow M;~P\longleftrightarrow -\Omega ;~V\longleftrightarrow J. \label{dic}
\end{eqnarray}

Now we are in a position to use the known tools of thermodynamics to black holes. In order to follow a Gibbsian approach to exhibit and classify black hole phase transitions it is customary to define the Gibb's free energy. Gibb's free energy and temperature of this black hole was calculated earlier in \cite{Cald, modak} and these are given by %First of all we plot the entropy-temperature graph for the Kerr-AdS black hole. The Hawking temperature for this black hole was calculated in \cite{other,modak} 
\begin{eqnarray}
G (S,\Omega) = M-TS-\Omega J \\ 
T (S,\Omega) = \left(\frac{S(\pi+S)^3}{\pi+S-S\Omega^2}\right)^{\frac{1}{2}}\times\nonumber\\
 \frac{\pi^2-2\pi S(\Omega^2-2)-3S^2(\Omega^2-1)}{4\pi^{\frac{3}{2}}S(\pi+S)^2}.
\label{temp}
\end{eqnarray}
 The relevant $S-T$ and $G-T$ plots are shown in fig.2. 
\begin{figure}[t]
\centering
%\rotatebox{270}{
\includegraphics[angle=0,width=8cm,height=4cm]{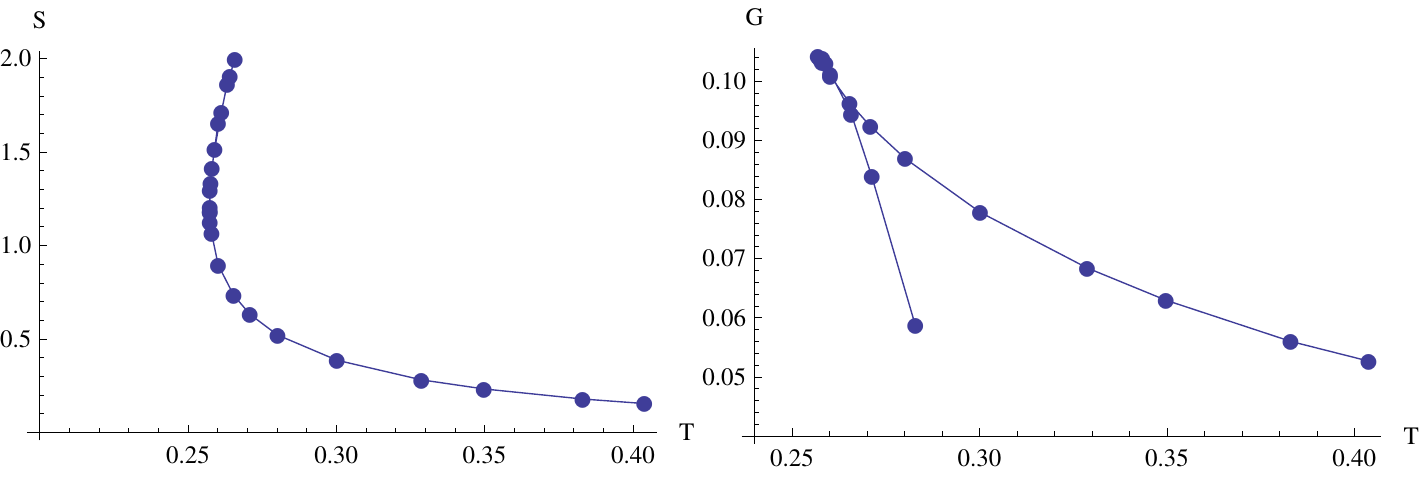}
\caption[]{Entropy ($S$) and Gibb's free energy ($G$) plot with respect to temperature ($T$) for fixed $\Omega=0.5$}
\label{figure 2a}
\end{figure}
The temperature-entropy plot is continuous. Also, there exists a minimum temperature ($T_c=0.2574$) after which the slope of this curve tends to change its sign very slowly. These features are qualitatively similar to fluids (or magnetic systems) \cite{stanely} revealing that there is no first order phase transition involving latent heat. Also, the same analogy indicates the onset of a continuous (higher order) phase transition with a critical temperature $T_c$. Contrary to what happens in usual fluids, however, the transition appears smeared instead of being sharp. Similar conclusions are also drawn from the $G-T$ plot.

The situation becomes more transparent when, once again following the analogous treatment for fluids, we plot the specific heat at constant angular velocity ($C_{\Omega}$) (analogue of $C_P$ for fluids) with entropy. The expression of $C_{\Omega}$ is given by \cite{modak},
%\begin{widetext}
\begin{eqnarray}
C_{\Omega}= 2S(\pi+S)(\pi+S-S\Omega^2)\times\nonumber\\
\frac{\pi^2-2\pi S(\Omega^2-2)-3S^2(\Omega^2-1)}{(\pi+S)^3(3S-\pi)-6S^2(\pi+S)^2\Omega^2+S^3(4\pi+3S)\Omega^4}\nonumber\\
\label{sph}
\end{eqnarray}
%\end{widetext}
The $C_{\Omega}-S$ plot (figure 3) is a concrete evidence of onset of a phase transition.  Now if we want to compare this plot with usual higher order phase transition we find one important difference. This is the appearance of a smeared region on either side of the critical point.

Let us just briefly discuss how the effects manifested in the curves of fig.2 are concretised in fig.3, which is quite akin to what happens for fluids, albeit with some crucial difference. First, the smeared nature of the phase transition is prominently displayed. Second, it is seen that the curvature of the two arms in fig.2 ($G-T$) change sign. Since the curvature of $G$ is just the specific heat, we observe the occurence of the phase transition curves in the positive and negative quadrant of the $C_{\Omega}-S$ plot seperated by the critical point. 

Clearly a Gibbsian approach, which has its root in mean field theory, breaks down in order to analyse this infinitely diverging, smeared phase transition. In the following we make a modification to known Gibbsian approach (Ehrenfest's scheme \cite{book}), incorporating a smearing effect, to analyse this new phase transition. We shall build our machinary considering a fluid system which undergoes a smeared phase transition and finally using (\ref{dic}) we shall recast relevant quantities for black holes.

%Let us recall the definition of Gibbs free energy
%\begin{eqnarray}
%G=E-TS+PV
%\label{equation 19a}
%\end{eqnarray}
%Infinitesimal variation of the above equation together with the basic thermodynamical law (\ref{fl})
%\begin{eqnarray}
%dE=TdS - PdV
%\label{de}
%\end{eqnarray}
 %gives
%\begin{eqnarray}
%dG=VdP-SdT
%\label{dg}
%\end{eqnarray}
%Clearly entropy and volume are first order derivatives of the Gibbs free energy, i.e. $S=-\left(\frac{\partial G}{\partial T}\right)_{P};~V= \left(\frac{\partial G}{\partial P}\right)_{T}$. Since one can in general interchange the order of derivatives
%\begin{eqnarray}
%\left(\frac{\partial }{\partial T}\left(\frac{\partial G}{\partial P}\right)_{T}\right)_{P}=\left(\frac{\partial }{\partial P}\left(\frac{\partial G}{\partial T}\right)_{P}\right)_{T}
%\label{max}
%\end{eqnarray}
%it yields the following Maxwell relation 
%\begin{eqnarray}
%\left(\frac{\partial V}{\partial T}\right)_{P}=-\left(\frac{\partial S}{\partial P}\right)_{T}
%\label{maxwellrelation}
%\end{eqnarray}

Now to generalize Ehrenfest's scheme we start from the fact that in a smeared phase transition the values of entropy or volume of two phases at the phase transition point are not exactly same as it should be in a second order phase transition, instead there are small smearing terms ($s,v$) in the following manner
\begin{subequations}
\begin{eqnarray}
%&&G_2=G_1+g\label{freeenergyequality}\\
&&S_2=S_1+s \label{entropyequality}\\
&&V_2=V_1+v\label{volumeequality}
\end{eqnarray}
\end{subequations}
where the subscripts 1 and 2 denote the values in the two phases. If we characterise the phase transition by the temperature ($T$) and pressure ($P$) for which (\ref{entropyequality}) and (\ref{volumeequality}) hold, at a different phase transition point (characterised by $T'=T+dT,P'=P+dP$) the following equations will be true
\begin{subequations}
\begin{eqnarray}
&&S_2+dS_2=S_1+dS_1+s'\label{2ndentropyequality}\\
&&V_2+dV_2=V_1+dV_1+v'\label{2ndvolumeequality}
\end{eqnarray}
\end{subequations}
From (\ref{entropyequality}) and (\ref{2ndentropyequality}) we find
\begin{eqnarray}
dS_1=dS_2+(s-s')\label{entropychange}
\end{eqnarray}
Taking $S$ as a function of $T$ and $P$
we write the infinitesimal change in entropy as $dS=\left(\frac{\partial S}{\partial T}\right)_{P}dT+\left(\frac{\partial S}{\partial P}\right)_{T}dP$. Using the Maxwell relation $\left(\frac{\partial V}{\partial T}\right)_{P}=-\left(\frac{\partial S}{\partial P}\right)_{T}$, this equation takes the form,
\begin{eqnarray}
dS &=& \frac{C_{P}}{T}dT-\chi dP,\label{dS}\\{\textrm{where,}}\nonumber\\
%\end{eqnarray}
%~{\textrm{where}} 
%\begin{equation}
\chi &=& \left(\frac{\partial V}{\partial T}\right)_{P};~C_{P} = T\left(\frac{\partial S}{\partial T}\right)_{P}\label{corsp}
\end{eqnarray}
are, respectively, the volume expansion index and the specific heat at constant pressure. Since (\ref{dS}) is independently true for two phases we write
%\begin{subequations}
\begin{eqnarray}
dS_1=\frac{C_{P_1}}{T}dT-\chi_1 dP;~~dS_2=\frac{C_{P_2}}{T}dT-\chi_2 dP \label{ds2}.
\end{eqnarray} 
%\end{subequations}
%Use of the condition (\ref{entropychange}) requires the equality of (\ref{ds1}) and (\ref{ds2}).
%\begin{eqnarray}
%\frac{C_{P_1}}{T}dT-\chi_1 dP=\frac{C_{P_2}}{T}dT-\chi_2 dP +(s-s').
%\end{eqnarray} 
Now substituting this in (\ref{entropychange}) we finally get the generalized first Ehrenfest's equation  
\begin{eqnarray}
\left(\frac{\partial P}{\partial T}\right)_{S}=\frac{C_{P_2}-C_{P_1}}{T(\chi_2-\chi_1)}+\frac{s-s'}{(\chi_2-\chi_1)(T'-T)}.
\label{eren1}
\end{eqnarray} 
In the absence of any smearing in entropy, the second term on the right hand side is zero and one recovers the first Ehrenfest's equation in its usual form.

Next considering (\ref{volumeequality}) and (\ref{2ndvolumeequality}) we find
\begin{eqnarray}
dV_1=dV_2+(v-v').\label{angularmomentumchange}
\end{eqnarray}
Taking $V$ as a function of $T$ and $P$, we find,
\begin{eqnarray}
dV&=&\left(\frac{\partial V}{\partial T}\right)_{P}dT+\left(\frac{\partial V}{\partial P}\right)_TdP\\
&=&\chi dT+\xi dP
\end{eqnarray}
where $\xi$ is the compressibility 
\begin{equation}
\xi=\left(\frac{\partial V}{\partial P}\right)_T
\label{compr}
\end{equation}
and $\chi$ has been defined in (\ref{corsp}). Following the previous steps we find the generalized second Ehrenfest's equation
\begin{eqnarray}
\left(\frac{\partial P}{\partial T}\right)_{V}=\frac{\chi_{2}-\chi_{1}}{\xi_{2}-\xi_{1}}+\frac{(v-v')}{(\xi_2-\xi_1)(T'-T)}.
\label{eren2}
\end{eqnarray} 
We can expect that in a smeared phase transition, (\ref{eren1}) and (\ref{eren2}) will be satisfied. In the following part of our paper we shall investigate this in the context of black hole phase transition.

Using (\ref{dic}), (\ref{eren1}) and (\ref{eren2}) we now propose the following generalized Ehrenfest's equation for the Kerr-AdS black hole,
\begin{eqnarray}
&& -\left(\frac{\partial \Omega}{\partial T}\right)_{S} = \frac{C_{\Omega_2}-C_{\Omega_1}}{T(\chi_2-\chi_1)}+\frac{s-s'}{(\chi_2-\chi_1)(T'-T)}\nonumber\\
\label{erenbh1}\\
&&-\left(\frac{\partial \Omega}{\partial T}\right)_{J} = \frac{\chi_{2}-\chi_{1}}{\xi_{2}-\xi_{1}}+\frac{(j-j')}{(\xi_2-\xi_1)(T'-T)}.\label{erenbh2}
\end{eqnarray} 

In order to check the validity of (\ref{erenbh1}) and (\ref{erenbh2}) we first recall the following equations from \cite{modak}
%\begin{widetext}
\begin{subequations}
\begin{eqnarray}
&&\chi = \nonumber\\ &&\frac{6S^2(\pi+S)^3\Omega-2S^3(\pi+S)(2\pi+3S)\Omega^3}{(\pi+S)^3(3S-\pi)-6S^2(\pi+S)^2\Omega^2+S^3(4\pi+3S)\Omega^4}\nonumber\\\label{25}
\end{eqnarray}
\begin{eqnarray}
\xi &=& S\sqrt{\frac{S(\pi+S)^3}{\pi+S-S\Omega^2}}F\label{JKT}\\
-\left(\frac{\partial\Omega}{\partial T}\right)_S &=& -\frac{4\pi^{\frac{3}{2}}(\pi+S-S\Omega^2)^2\sqrt\frac{S(\pi+S)^3}{(\pi+S-S\Omega^2)}}{S\Omega\left(S(\pi+S)(2\pi+3S)\Omega^2-3(\pi+S)^3\right)}
\nonumber\\\label{31}\\
-\left(\frac{\partial\Omega}{\partial T}\right)_J &=& \sqrt{\frac{\pi+S-S\Omega^2}{S(\pi+S)^3}}H\label{32}
\end{eqnarray}
\end{subequations}
where $F$ and $H$ are defined as {\footnote{$F=\frac{(\pi-3S)(\pi+S)^3-2S(\pi+S)^2(4\pi+3S)\Omega^2+S^2(2\pi+3S)^2\Omega^4}{2\pi^{\frac{3}{2}}[(\pi-3S)(\pi+S)^4+6S^2(\pi+S)^3\Omega^2-S^3(\pi+S)(4\pi+3S)\Omega^4]}\\
H=\frac{4\pi^{3/2}S(\pi+S)^2\Omega(\left(3(\pi+S)^2-S(2\pi+3S)\Omega^2\right))}{(\pi+S)^3(3S-\pi)+2S(\pi+S)^2(4\pi+3S)\Omega^2-S^2(2\pi+3S)^2\Omega^4}$}}.
%\end{widetext}
\begin{table*}[th]
\centering
\squeezetable
\begin{tabular}{|c|c|c|c|c|c|c|c|c|}
\hline
 &\multicolumn{4}{c|}{1st generalized Ehrenfest's reln.} &\multicolumn{3}{|c|}{2nd generalized Ehrenfest's reln.} & 
%&\multicolumn{2}{c|}{EP($\leftarrow$)}
\\
\cline{2-9}
 $\Omega$&l.h.s$=-\left(\frac{\partial\Omega}{\partial T}\right)_{J}$&$X_1$ &$Y_1$&r.h.s=$X_1+Y_1$&l.h.s$=-\left(\frac{\partial\Omega}{\partial T}\right)_{S}$&$X_2$& $Y_2$&r.h.s=$X_2+Y_2$ 
\\
\hline\hline
    0.1 & 72.1469 & 64.6853 & 0.0 & 64.6853 & 66.8405 & 6.232  & 39.50 & 45.732\\ \hline
    0.2 & 35.6305 & 31.6156 & Do  & 31.6156 & 35.5666 & 9.470  & 20.56 & 30.030\\ \hline
    0.3 & 23.2151 & 20.5940 & Do  & 20.5940 & 23.2601 & 10.329 & 12.36 & 22.689\\ \hline
    0.4 & 16.8036 & 15.0467 & Do  & 15.0467 & 16.7774 & 9.833  & 5.13  & 14.963\\ \hline
    0.5 & 12.8027 & 11.9817 & Do  & 11.9817 & 12.7985 & 8.578  & 2.924  & 11.502\\ \hline
    0.6 & 9.9750 & 8.9837   & Do  & 8.9837 & 9.9852   & 7.484  & 1.26  & 8.744\\ \hline
    0.7 & 7.7547 & 7.1375   & Do  & 7.1375 & 7.7544   & 5.944  & 0.596 & 6.540\\ \hline
    0.8 & 5.8653 & 5.4616   &Do   & 5.4616 & 5.8674   & 5.025  & 0.262 & 5.287\\ \hline
    0.9 & 4.0034 & 3.3300   & Do  & 3.3300 & 4.0030   & 2.970  & 0.188 & 3.158\\ \hline
   % 0.99 & 1.5252 & 1.1814  & Do  & 1.1814 & 1.5254   & 1.090  & 0.015 & 1.11\\ \hline
\end{tabular} \caption{Summary of results for validity of generalized Ehrenfest's relations for Kerr-AdS black hole for various angular velocities.}
\end{table*}
The physically allowed values for $\Omega$ are $0\le\Omega\le 1$, else the Hawking temperature becomes negative \cite{modak}. We choose a particular value around the central range ($\Omega=0.5$) to examine how this generalized scheme works. For this value of $\Omega$ the plots of $C_{\Omega}$, $\chi$ and $\xi$ (see figures 3 and 4) show a smeared divergence at the critical entropy ($S_{c}=1.208$). For these values of $\Omega$ and $S$ the critical temperature is found to be $T_c=0.2574$ from (\ref{temp}).
\begin{figure}[ht]
%\centering
%\rotatebox{270}{
\includegraphics[angle=0,width=8cm,height=3.5cm]{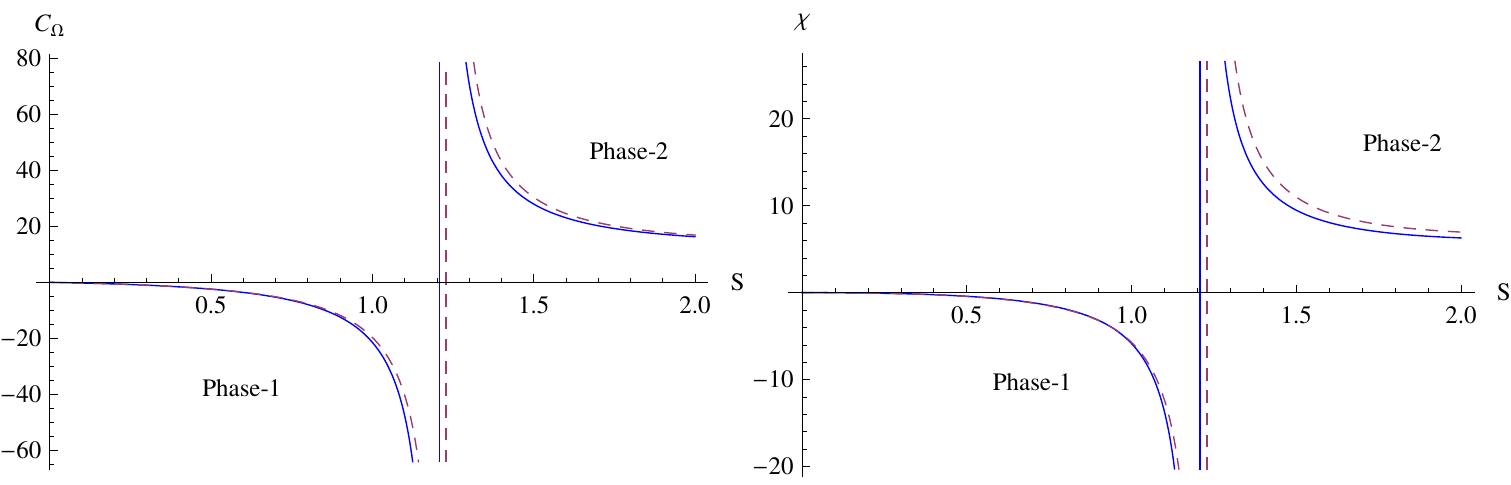}
\caption[]{{Semi-classical Specific heat ($C_{\Omega}$) and volume expansion index ($\chi$) vs entropy ($S$) plot. The solid line stands for $\Omega=0.5$ and the dashed line corresponds to its infintely small variation ($\Omega'=\Omega + 5\%\times\Omega$).}}
\label{figure 1}
\end{figure}
%\begin{figure}[h]
%\centering
%\rotatebox{270}{
%\includegraphics[angle=0,width=6cm,height=3cm]{chi.pdf}
%\caption[]{{Change in angular momentum coefficient ($\chi$) vs entropy ($S$) plot for fixed $\Omega=0.5$.}}
%\label{figure 2}
%\end{figure}
\begin{figure}[h]
\centering
%\rotatebox{270}{
\includegraphics[angle=0,width=8cm,height=3.5cm]{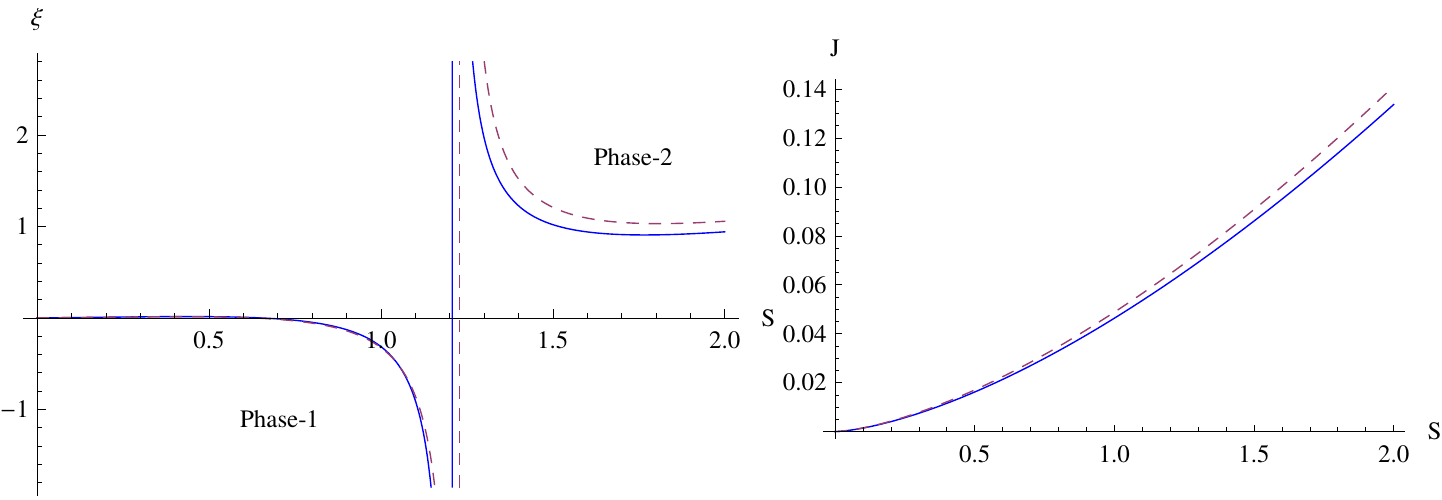}
\caption[]{{Moment of inertia index ($\xi$) and angular momentum ($J$) vs entropy ($S$) plot. The solid line stands for $\Omega=0.5$ and the dashed line corresponds to its infintely small variation ($\Omega'=\Omega + 5\%\times\Omega$).}}
\label{figure 3}
\end{figure}

Now the left hand sides of (\ref{erenbh1}), (\ref{erenbh2}) evaluated by using (\ref{31}), (\ref{32}) at the critical point ($S_c=1.208, T_c=0.2574$ for $\Omega=0.5$) are found to be $12.8027$ and $12.7925$. To calculate the right hand side we first draw the $C_{\Omega}-S,~\chi -S.~\xi -S$ and $J-S$ plots (figures 3,4) using (\ref{sph}), (\ref{25}), (\ref{JKT}) and (\ref{JJ}), respectively. Let us first consider the smearing independent terms in (\ref{erenbh1}), (\ref{erenbh2}). The respective values of relevant quantities are borrowed from their plots. We choose such points in two phases (say $S_1$ for phase 1 and $S_2$ for phase 2) by avoiding the smearing region as in this part values widely change for two neighbouring points. Our prescription of choosing a point is that it should be very close to the critical point and belong to a region where the slope of the curve is vanishingly small. For $\Omega=0.5$ we take $S_1=0.653$ and $S_2=1.763$. Thus we have a smeared region of width $s=S_2-S_1=1.11$ (see (\ref{entropyequality})). For this set of values the changes in $C_{\Omega}, \chi, \xi$ between the two phases are found to be $\Delta C_{\Omega}=23.80~, \Delta\chi=7.72,~\Delta\xi=0.90$. The values thus found for the first terms of the right hand side of (\ref{erenbh1}) and (\ref{erenbh2}) are $11.9817$ and $8.578$ respectively. We now observe that, since the l.h.s. of (\ref{erenbh1}) equals 12.8027, the r.h.s. almost matches. This does not hold for (\ref{erenbh2}). This is reminiscent of recent studies \cite{Banerjee} which show that if one neglects the smearing effect for black hole phase transitions then only the first Ehrenfest's equation is satisfied and the second one is violated. Effectively, therefore the smearing terms have a nontrivial contribution only for (\ref{erenbh2}). We now discuss this issue.

In order to obtain the second (smearing dependent) terms we first make a small change ($5\%$) in angular velocity and see the behaviour of all the plots. Let us first consider (\ref{erenbh1}). It is clear from all plots that the smearing width suffers no significant change, i.e. $s-s'\approx 0$. We may neglect this small change as it is further weakend by the denominator of $(s-s')$ in (\ref{erenbh1}). As a result the right hand side of the first equation almost remains unchanged. Regarding (\ref{erenbh2}), however, this small change in $\Omega$ changes the smearing in $J$. It is obtained by finding $j (=J(S_2=1.763,\Omega=0.5)-J(S_1=0.653,\Omega=0.5))$ and $j' (=J(S_2=1.763,\Omega=0.525)-J(S_1=0.653,\Omega=0.525))$ from (\ref{JJ}) and then taking their difference. This value of $j-j'$ in (\ref{erenbh2}) is comparable with its denominator. Thus it is no longer negligible. The contribution from this extra term is found to be $\frac{j-j'}{(\xi_2-\xi_1)(T'-T)}=\frac{(0.091-0.086)}{0.0019\times0.9}=2.924$. This value when added with the other term (8.578) gives the right hand side of (\ref{erenbh2}) as 11.502. This is in reasonable agreement with the l.h.s. value (12.7925). In order to get a full picture we consider various allowed values of $\Omega$ and check the validity of (\ref{erenbh1}) and (\ref{erenbh2}). The results are summarised in a tabular form (see Table-I \footnote{The new variables in table-II are defined as $X_1=\frac{C_{\Omega_2}-C_{\Omega_1}}{T_c(\chi_2-\chi_1)}, Y_1=\frac{s-s'}{(T'_c-T_c)(\chi_2-\chi_1)},~X_2=\frac{\chi_2-\chi_1}{\xi_2-\xi_1},~Y_2=\frac{j-j'}{(T'_c-T_c)(\xi_2-\xi_1)}$}). It clearly shows the validity of these new equations for black hole phase transition.

We have discussed a new approach to study phase transitions in black holes that is based on their analogy with fluids. Several features of this analogy played a crucial role. Unlike the liquid to vapour transition, here it is not first order. Rather, according to our analysis this is a smeared continuous (higher order) phase transition. Following a generalized Gibbsian approach we have derived the governing equations for such transitions. These equations were verified for phase transition in Kerr-AdS black hole. Our calculations are robust. We checked this by taking other values of $S_1$ and $S_2$ close to those taken here. We feel that the approach discussed here can be pushed further, by exploiting the analogy with fluids, to gain a deeper insight into black hole thermodynamics.  \\  

\noindent One of the authors S. K. M thanks the Council of Scientific and Industrial Research (C. S. I. R), Government of India, for financial help. 
%\end{widetext}
%%%%%%%%%%%%%%%%%%%%%%%%%%%%%%%%%%%%%%%%%%%%%%%%%%%%%%%%%%%%%%%

%%%%%%%%%%%%%%%%%%%%%%%%%%%%%%%%%%%%%%%%%%%%%%%%%%%%%%%%%%%%%%%%%%%%%%%%%%%%%%%%
\end{document}